\documentclass[preprint,showpacs,preprintnumbers,amsmath,amssymb]{revtex4}

\usepackage{graphicx}

\begin{document}

\title{
Possible altitudinal, latitudinal and directional dependence of 
relativistic Sagnac effect in Chern-Simons modified gravity 
}
\author{Daiki Kikuchi}
\author{Naoya Omoto}
\author{Kei Yamada}
\author{Hideki Asada} 
\affiliation{
Faculty of Science and Technology, Hirosaki University,
Hirosaki 036-8561, Japan} 

\date{\today}

\begin{abstract}
Toward a test of parity violation in a gravity theory, 
possible effects of Chern-Simons (CS) gravity 
on an interferometer have been recently discussed. 
Continuing work initiated in an earlier publication 
[Okawara, Yamada and Asada, Phys. Rev. Lett. 109, 231101 (2012)], 
we study possible altitudinal and directional dependence of relativistic 
Sagnac effect in CS modified gravity. 
We compare the CS effects on Sagnac interferometers with 
the general relativistic Lense-Thirring (LT) effects. 
Numerical calculations show that the eastbound Sagnac interferometer 
might be preferred for testing CS separately, 
because LT effects on this interferometer cancel out. 
The size of the phase shift induced in the CS model 
might have an oscillatory dependence also on the altitude of the interferometer 
through the CS mass parameter $m_{CS}$. 
Therefore, the international space station site 
as well as a ground-based experiment is also discussed. 
\end{abstract}


\maketitle

\section{Introduction}
Modifications of the theory of general relativity (GR) have been of interest. 
Particularly, some modifications that introduce second (or higher) 
order terms of curvature tensors represent high-energy corrections 
to the Einstein-Hilbert action. 
The Chern-Simons (CS) correction is one of modified gravity models. 
The CS modification is not an {\it ad hoc} extension, 
but it is actually motivated by both string theory, as a
necessary anomaly-canceling term to conserve unitarity
\cite{Polchinski}, 
and 
loop quantum gravity (LQG), as 
a counter term for the anomaly\cite{Ashtekar} 
and recently as the emergence of the CS gravity when 
the Barbero-Immirzi parameter of LQG is promoted to 
a scalar field and the Holst action is coupled to fermions 
\cite{Taveras}. 
CS modifications to gravity were first formulated 
in 2+1 dimensions \cite{Deser}. 
Several authors investigated the structure of these theories in 3+1 
dimensions to show that they could arise as a low-energy limit 
of string theory \cite{Kaloper}. 
The theory and formulation of CS modified gravity have been discussed 
in a number of papers (see \cite{AY2009} for a review), 
and possible imprint of such a modification 
in the early universe has been recently investigated. 
Moreover, there has been little work on tests of 
such CS corrections in the present Universe.

In nondynamical CS gravity, 
a scalar field is assumed to be externally prescribed. 
It is often taken to be a linear function of the coordinate time 
(as a canonical choice), and induces parity violation in the theory. 
Nondynamical CS gravity depends on a single free parameter 
\cite{AY1, AY2, Smith}. 
The constraint on this parameter with measurements of 
frame-dragging of bodies orbiting the Earth has been discussed; 
The proposal has been implemented by Ali-Haimoud and Chen 
\cite{Ali-Haimoud} to constrain CS gravity;  
Yunes and Spergel, and Ali-Haimoud \cite{YS, Ali2011} have used 
double-binary-pulsar measurements. 

In addition to cosmological and astrophysical tests, 
current attempts to probe general relativistic effects 
in quantum mechanics focus on precision measurements of phase shifts 
in quantum interferometers (e.g. \cite{Zych}). 
Toward a test of parity violation in a gravity theory beyond GR, 
Okawara and his collaborators have recently studied 
a possible constraint by neutron interferometers 
\cite{OYA, OYA2}, where they used Alexander and Yunes (AY) model. 
The main purpose of the present paper is to improve the previous results 
on the interferometers regarding two points \cite{OYA,OYA2}. 
One improvement is that the present paper considers 
an up-dated nondynamical CS model that has been developed 
by Smith, Erickcek, Caldwell, and  Kamionkowski (SECK) \cite{Smith} 
in order to study both interior and exterior 
gravitational fields by a spinning object, 
whereas AY model assumes a point-like spinning object. 
SECK model can treat an extended source of the gravitational field and, 
in some limit, it approaches AY model. 
Because of including a mass parameter 
(through a homogeneous solution to the field equation), 
SECK model shows oscillating behavior of 
the gravitational potential 
along the radial direction of a central object. 
As a result, we shall study altitudinal dependence of Sagnac effect 
in the present paper. 
The other improvement is that we consider Sagnac interferometers in optics. 
This is more advantageous at present, 
mainly because it is relatively easy to put at different places 
Sagnac interferometers compared with neutron interferometers 
that need nuclear reactors as a source of neutrons.

This paper is organized as follows. 
Sec. II briefly reviews SECK model of nondynamical CS gravity theory 
and the relativistic Sagnac effect. 
In Sec. III, we compute relativistic Sagnac effects in the CS model.  
Sec. IV provides numerical calculations. 
Sec. V is devoted to conclusion. 

Throughout this paper, Latin indices run from $1$ to $3$, 
while Greek ones from $0$ to $3$.

\section{Relativistic Sagnac effect and nondynamical CS modified gravity} 
This section summarizes the basics of computing 
a phase difference in Sagnac interferometer 
by CS modified gravity.  

\subsection{Relativistic Sagnac effect} 
The Sagnac effect, which is often called Sagnac interference, 
originally describes a phenomenon encountered in interferometry 
that is elicited by rotation. 
It appears manifestly in a setup called a ring interferometer. 
Similar effects due to relativistic gravitomagnetic fields 
in a stationary spacetime are often called relativistic Sagnac effects. 
For instance, see \cite{Stedman,SW} for a review of 
ring-laser tests of fundamental physics. 
Yet, a recent proposal of an experimantal scheme to measure 
the Lense-Thirring (LT) effect with a Sagnac interferometer 
is still a long way from reality (See e.g. \cite{Tartaglia}). 
Accroding to \cite{Tartaglia}, 
{\it G in Geodetic Observatory Wettzell}, 
the best ring laser in the world, already achieves 
the accuracy within one order of magnitude from the expected signal 
for detections of LT effects. 
Aiming at the LT detection, they are now planning a new experiment 
called {\it GINGER} in order to reduce various sources of noises. 
Therefore, such a drastic experimental progress is a major premise 
for our ambitious proposal of using a Sagnac interferometer 
in order to testify to the CS effect separately from 
the LT one (that has not been detected with any interferometer yet).

Consider two beams of monochromatic light in a closed path 
(denoted by $C$) 
such as a ring or a square, where one beam is clockwise and the other 
anticlockwise. 
Along the light path, we have 
$ds^2 = g_{\mu\nu} dx^{\mu} dx^{\nu} = 0$. 
In the gravitomagnetic field, 
the leading contribution to the arrival time shift $\Delta t$ 
is given by the relativistic version of Sagnac effect as 
\cite{Anandan} 
\begin{align}
c \Delta t = - 2 \oint_C \frac{g_{0i}} {g_{00}} dx^i , 
\label{Delta1}
\end{align}
where 
$C$ denotes a 
{\bf
clockwise 
} 
closed path of a light beam. 
This formula is almost the same as that for a matter wave 
\cite{OYA, OYA2} except for a factor 2, 
if de Broglie wavelength is replaced by the photon wavelength.
See \cite{Rizzi,Zendri} for relativistic higher order corrections. 
Dividing the time shift by the wavelength of a photon $\lambda$, 
we obtain the phase difference as 
\begin{align}
\Delta \Phi = \frac {2 \pi } {\lambda} c \Delta t
\label{DeltaPhi}
\end{align}

Let us consider experiments near the surface of Earth, 
for which we can assume a small perturbation around the Minkowskian 
background spacetime as $g_{\mu\nu} = \eta_{\mu\nu} + h_{\mu\nu}$. 
The time-space component of the metric 
does matter in the relativistic Sagnac formula. 
It is denoted as a spatial vector 
$\vec{h} \equiv (h_{01}, h_{02}, h_{03})$. 
The leading order of Eq. (\ref{Delta1}) becomes 
\cite{Anandan} 
\begin{align}
c \Delta t = - 2 \int_S (\vec{\nabla} \times \vec{h}) \cdot d\vec{S} + O(h^2) , 
\label{Delta2}
\end{align}
where we used Stokes theorem, 
$d\vec{S}$ denotes the infinitesimal areal vector, 
and $S$ means the area of the Sagnac interferometer. 

Note that the relativistic Sagnac effect is dependent on 
the inner product as $(\vec{\nabla} \times \vec{h}) \cdot \vec{N}_I$ 
for the unit normal vector $\vec{N}_I$ to the interferometer plane, 
whereas the relativistic gyroscope precession by $\vec{h}$ 
depends mainly on the outer product as 
$(\vec{\nabla} \times \vec{h}) \times \vec{L}$ 
for the spin vector $\vec{L}$, roughly speaking.

\subsection{CS modified gravity}
Following Ref. \cite{Smith}, we consider a CS modification 
to general relativity. 
The present paper focuses on the leading-order CS correction 
due to the rotation of a central body. 
Here, the Earth is approximated by a spinning body 
that is a source of the gravitational field.  
Therefore, it is sufficient to consider 
nondynamical CS gravity in this paper, 
though more dynamical systems such as compact binaries and 
black hole formations may require a dynamical CS treatment  
because of their rapid changes in time and space \cite{note}.

The exterior weak-field $\vec{h}$ in GR, 
which causes the LT effect, 
is known to be \cite{Smith}
\begin{align}
\vec{h}_{LT} = \frac {4 G M R^2} {5 c^3 r^2} 
\left( \vec{n} \times \vec{\omega} \right) , 
\label{g0i_GR}
\end{align}
where $R$ is the radius of Earth, $M$ is its mass, 
$\vec{\omega}$ is its angular velocity, 
$r$ is the distance from the origin, 
and $\vec{n}$ is the unit vertical vector.

We consider the the CS modified gravity theory 
by the action as \cite{Smith} 
\begin{eqnarray}
S = \int d^4x \sqrt{-g} 
\left[ -\frac{c^4}{16 \pi G} \mbox{\boldmath $R$} 
+ \frac{\ell}{12} \theta \mbox{\boldmath $R$} \tilde{\mbox{\boldmath $R$}} 
- \frac12 (\partial \theta)^2 - V(\theta) + {\cal L}_{mat} \right] , 
\end{eqnarray}
where ${\cal L}_{mat}$ is the Lagrangian density for matter, 
$g \equiv \det{(g_{\mu\nu})}$, and $\mbox{\boldmath $R$}$ is the Ricci scalar, 
and $\mbox{\boldmath $R$} \tilde{\mbox{\boldmath $R$}}$ 
denotes a contraction of the Riemann tensor and its dual, 
and $\theta$ is a dynamical scalar field with a potential $V(\theta)$. 
In this theory, we follow Ref. \cite{Smith} 
to suppose that the scalar field depends only on cosmic time, 
$\theta = \theta(t)$, 
and define $m_{CS} \equiv -3/(\ell \kappa^2 \dot{\theta})$, 
where $\kappa = 8\pi G/c^4$. 
The spacetime metric as the weak-field solution 
to the CS modified field equations 
appears at the leading order in $g_{0i}$. 
It is obtained as \cite{Smith}
\begin{align}
\vec{h}_{CS} = \frac{12 G M} {m_{CS} c^3 R} \left[ C_1 (r) \vec{\omega} + 
C_2 (r) \vec{n} \times \vec{\omega} + C_3 (r) \vec{n} \times \left( \vec{n} \times \vec{\omega} \right) \right] , 
\label{g0i_CS}
\end{align}
with 
\begin{align}
C_1(r) &= \frac{2 R^3} {15r^3} + \frac{2R} {r} j_2(m_{CS} R) y_1(m_{CS}r), \notag \\
C_2(r) &= m_{CS} R j_2(m_{CS} R) y_1(m_{CS}r), \notag \\
C_3(r) &= \frac{R^3} {5r^3} + m_{CS} R j_2(m_{CS} R) y_2(m_{CS}r) , 
\end{align}
outside the sphere. 
Here, $j_{\ell}(x)$ and $y_{\ell}(x)$ are spherical Bessel functions 
of the first and second kind, respectively. 
Smith et al. obtained both the gravitomagnetic field 
inside and outside a spinning sphere. 
We focus on the exterior field, because interferometers are  
considered in this paper. 

Furthermore, curl of the field is obtained for GR as \cite{Smith} 
\begin{align}
(\vec{\nabla} \times \vec{h}_{LT} ) 
&= - \frac{ 4 G M R^2 } {5 c^3 r^3 } \left[ 
2 \vec{ \omega } + 3 \vec{n} \times \left( \vec{n} \times \vec{ \omega } \right)
\right] , 
\label{rot_GR}
\end{align}
For the CS term, it becomes \cite{Smith} 
\begin{align}
(\vec{\nabla} \times \vec{h}_{CS}) = 
- \frac{12 G M} {c^3 R} 
\left[ D_1 (r) \vec{ \omega } + D_2 (r) \vec{n} \times \vec{ \omega } 
+ D_3 (r) \vec{n} \times \left( \vec{n} \times \vec{ \omega } 
\right) \right] ,  
\label{rot_CS}
\end{align} 
with 
\begin{align}
D_1(r) &=\frac{2R} {r} j_2(m_{CS} R) y_1(m_{CS} r) , \\
D_2(r) &= m_{CS} R j_2(m_{CS} R) y_1(m_{CS} r) , \\
D_3(r) &= m_{CS} R j_2(m_{CS} R) y_2(m_{CS} r) . 
\end{align}

\section{Relativistic Sagnac effect induced by CS correction terms}

\subsection{Time shift and phase shift}
Substituting Eq. (\ref{rot_GR}) into Eq. (\ref{Delta1}) 
leads to the coordinate-time shift as 
\begin{align}
(c \Delta t)_{LT}  
&= \frac{8GMR^2} {5 c^3} 
\int_S \left[ \frac { 2 \vec{ \omega } + 3 \vec{n} \times \left( \vec{n} \times \vec{ \omega } \right) } {r^3} \right] 
\cdot \vec{N}_I dS \notag \\
&= \frac{8GM R^2 S} {5 c^3 r^3} 
\vec{N}_I \cdot
\left[ 2 \vec{ \omega } - 3 \vec{ \rho } \right]
\label{deltat-GR}
\end{align}
where we assumed that the size of the interferometer is much smaller 
than the radius of the Earth. 
Similarly, we obtain the time difference due to CS as 
\begin{align}
(c \Delta t)_{CS} 
&= \frac{24 G M} {c^3 R} 
\int_S \left[ D_1 (r) \vec{ \omega } + D_2 (r) \vec{n} \times \vec{ \omega } 
+ D_3 (r) \vec{n} \times \left( \vec{n} \times \vec{ \omega } \right) \right] 
\cdot \vec{N}_I dS \notag \\
&= \frac{24 G M S}{c^3 R} 
\vec{N}_I \cdot \left[ D_1(r) \vec{\omega} 
- D_2(r) \vec{\lambda} - D_3(r) \vec{\rho} \right] ,  
\label{deltat-CS}
\end{align} 
where $\vec{\lambda} \equiv \vec{\omega} \times \vec{n}$ 
is a vector parallel to a line of latitude on the sphere and 
$\vec{\rho} \equiv \vec{n} \times (\vec{\omega} \times \vec{n}) 
= \vec{\omega} - (\vec{\omega} \cdot \vec{n}) \vec{n}$ 
is a vector parallel to a line of longitude on the sphere. 


The order of magnitude of time difference by LT effects in GR is 
\begin{align}
(c \Delta t)_{LT} \sim \frac{8 G M S \omega} {5 c^3 R } . 
\end{align}
The magnitude of Eq. (\ref{deltat-CS}) is roughly  
\begin{align}
(c \Delta t)_{CS} \sim \frac{24 G M S \omega} {c^3 m_{CS} R^2} . 
\end{align}
From these equations, one can estimate the size of 
the relativistic Sagnac effect due to LT and CS, respectively.

The altitudinal dependence of CS effects 
might be negligible very near the surface of Earth. 
However, the difference between the ground-based interferometer 
and one at an altitude $\sim 400$ km (corresponding to the space station) 
might become of the order of the unity for 
$m_{CS} \sim 0.01 - 0.001 \mbox{km}^{-1}$, 
for instance. 
This point will be discussed later in more detail.

Here, we mention experimantal realities. 
With $m_{CS} = 0.001 \mbox{km}^{-1}$, 
one finds the shift $c \Delta t$ 
is of the order $10^{-18} \mbox{km}$ 
for an interferometer with an area of a square kilometer. 
In terms of strain $c \Delta t /L$, 
where $L$ is the size of the interferometer, 
the strain is $\sim 10^{-18}$. 
For a more modest meter-scale interferometer, 
the strain is three orders of magnitude smaller. 
While the modest setup might be more promising, 
the strain of $10^{-21}$ is apparently reachable 
by gravitational wave detecters such as LIGO (on the ground) 
and LISA (in the space). 
However, this is not the case. 
The strain under study is essentially a DC 
(namely zero-frequency) strain, 
while gravitational-wave interferometers 
search high-frequency signals 
and they try to kill various noises at low frequency.

\subsection{Numerical Calculations} 
According to previous works on the precession,  
\cite{AY1,AY2,Smith,Ali2011} 
there has been a constraint on $m_{CS}$ 
as $m_{CS} > 0.001 [\mbox{km}^{-1}]$, roughly speaking. 
Taking account of this existing constraint, 
numerical calculations are done for a parameter region 
$0.001 [\mbox{km}^{-1}] < m_{CS} < 0.1 [\mbox{km}^{-1}]$ in this paper. 
The time shift $(c \Delta t)_{CS}$ depends on four parameters 
as the CS mass parameter $m_{CS}$, 
the interferometer direction $\alpha$, 
the latitude $\phi$, 
and the altitude $h$, 
where $\alpha$ is defined as a horizontal angle measured 
clockwise from a north base line or meridian.  
For instance, $\alpha = 0^{\circ}$ and $90^{\circ}$ 
correspond to the direction along $\vec{\rho}$ and $\vec{\lambda}$, 
respectively.  
Furthermore, the shift is dependent also on the zenith angle, 
which is not considered in the present paper. 

\begin{figure}[t]
\includegraphics[width=14cm]{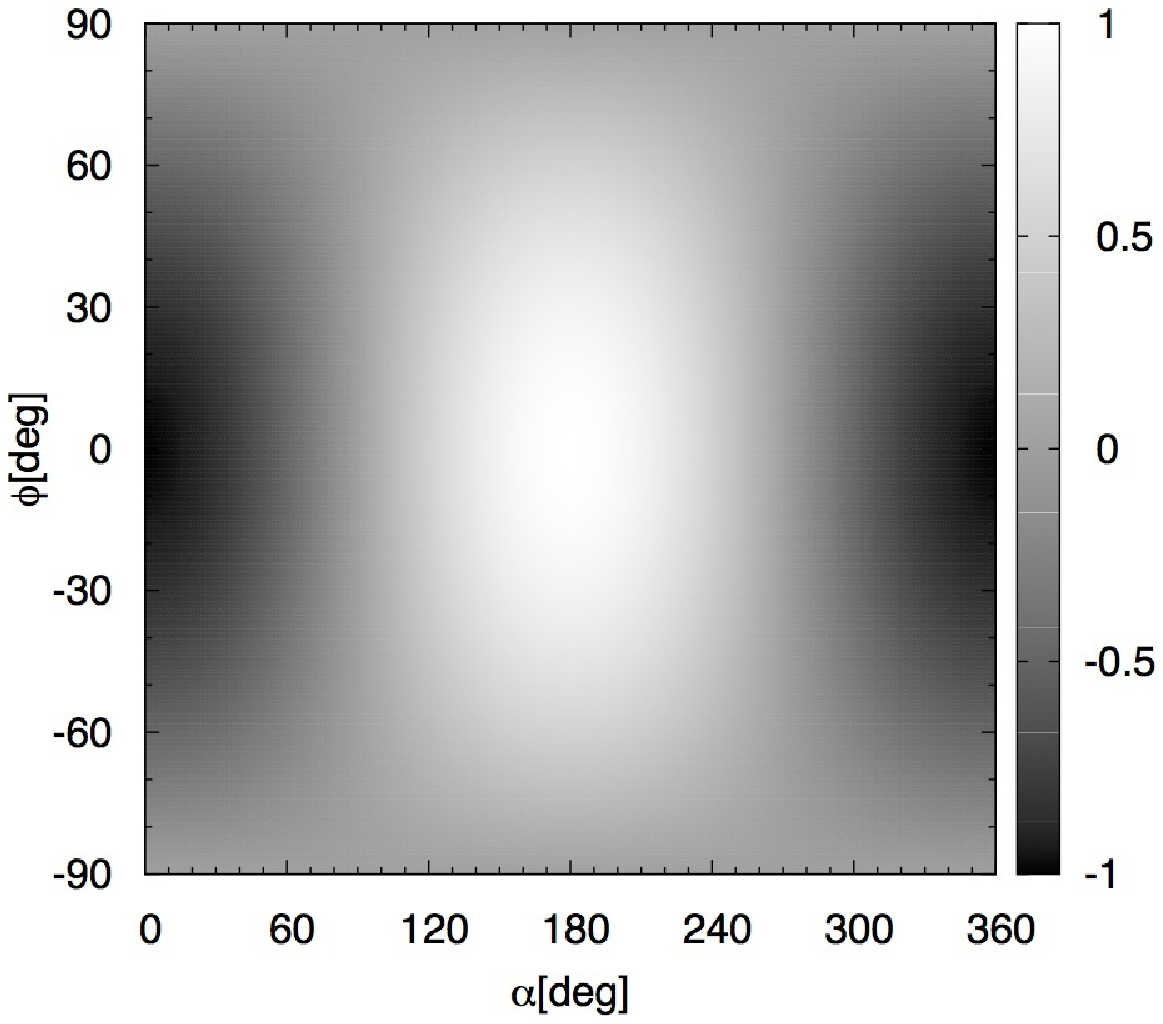}
\includegraphics[width=14cm]{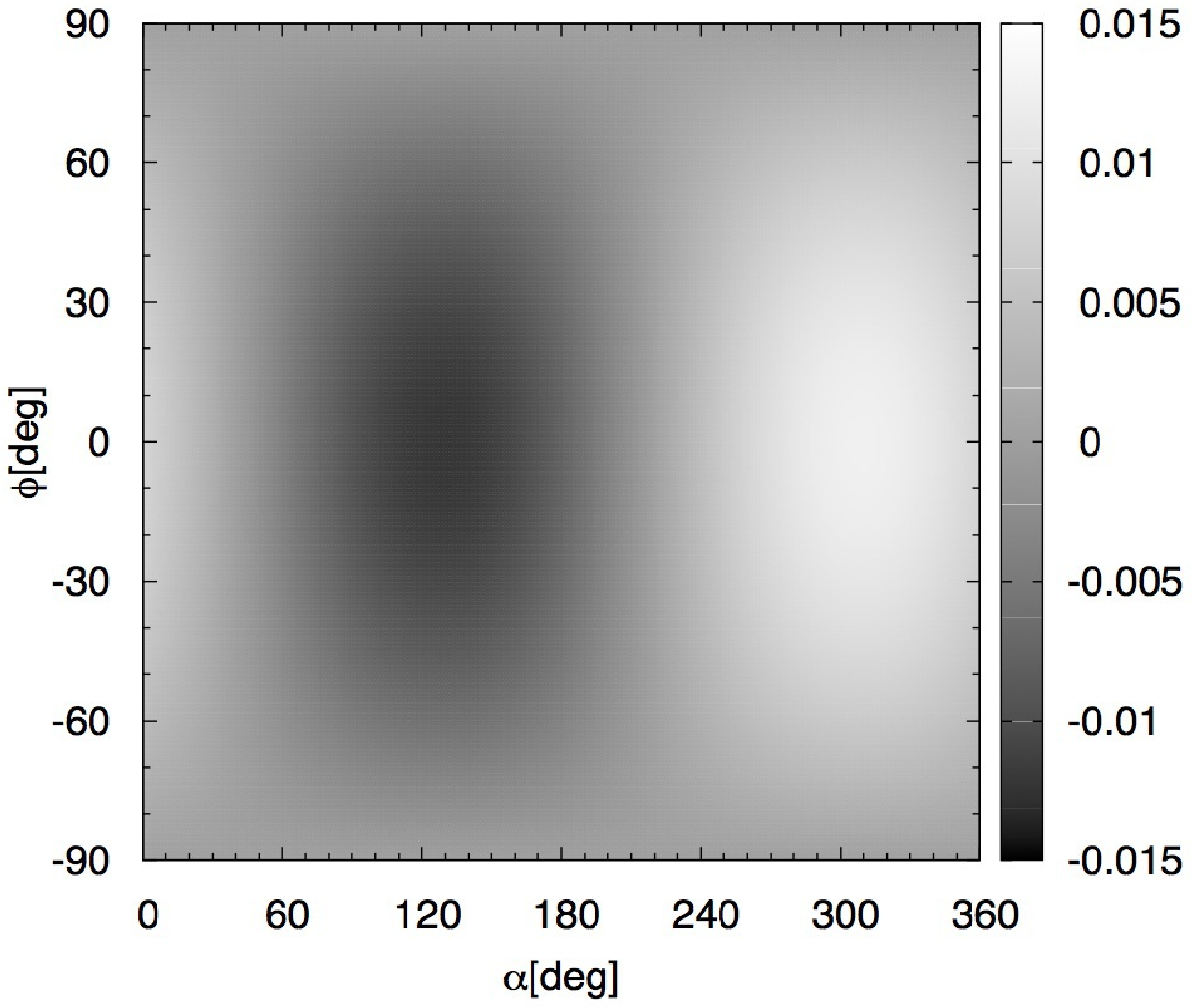}
\caption{ 
Contour maps for the dependence of time shift 
on the interferometer direction angle 
$\alpha$ and the latitude $\phi$, 
where the height corresponds to the angular part of $(c \Delta t)$. 
Top: $\Delta_{LT}$ by LT effects in GR. 
Bottom: $\Delta_{CS}$ by CS effects. 
}
\label{f1}
\end{figure}

As a reference, 
let us consider the time shift $(c \Delta t)_{LT}$ 
due to LT effects in GR. 
It is useful to define $\Delta_{LT}$ 
as $(c \Delta t)_{LT}$ in the units of $8 G M S \omega/5 c^3 R$, 
namely the angular part of $(c \Delta t)_{LT}$ 
as 
$\vec{N}_I \cdot 
\left[ 2 \vec{ \omega } - 3 \vec{ \rho } \right]/|\vec{\omega}|$. 
Figure \ref{f1} shows the dependence of $\Delta_{LT}$ 
on the latitude and the direction. 
It follows that there are no oscillating behaviors 
in LT effects. 
Note that LT effects vanish in the interferometer direction as 
$\alpha = 90^{\circ}$ and $270^{\circ}$ almost everywhere 
except for polar regions. 
This suggests that the eastbound direction of the interferometer 
might be preferred for testing CS, separately, 
because LT effects on this interferometer cancel out.

First, we consider ground-based experiments $(h=0)$, 
for which $(c \Delta t)_{CS}$ depends on the other three parameters 
$m_{CS}$, $\alpha$ and $\phi$. 
It is useful to define $\Delta_{CS}$ 
as $(c \Delta t)_{CS}$ in the units of $24 G M S \omega/c^3 R$, 
namely the angular part of $(c \Delta t)_{CS}$. 
Figure \ref{f1} shows numerical calculations of 
the dependence of the time shift $\Delta_{CS}$ 
on the latitude $\phi$ and the direction $\alpha$, 
where we assume $m_{CS} = 0.001 [\mbox{km}^{-1}]$ 
and the vertical axis denotes $\Delta_{CS}$. 
For this case, the CS effect becomes the largest 
around $\alpha = 130^{\circ}$ and $300^{\circ}$ 
for wide latitude regions from the equator to the middle latitude 
except for the polar regions. 
 

%
The latitudinal and directional dependence is weak around $m_{CS} \sim 0.1$, 
while it is strong around $m_{CS} \sim 0.001 - 0.01$. 
Therefore, one can say that the CS latitudinal and directional effects 
might be important in experiments, especially when we investigate 
the parameter region $m_{CS} \sim 0.001 - 0.01$. 
Even if any imprint by CS were marginally detected in the future 
(presumably at a low signal-to-noise ratio), 
it would be difficult to disentangle the CS signal from 
other effects without taking account of these dependencies. 
A comparison of phase measurements at two (or more) directions at 
different latitudes would be helpful for improving the CS bound or 
distinguishing the CS signal from others. 
Namely, a signal-to-noise ratio could be enhanced 
by a combined analysis of phase measurements 
at different latitudes and directions.

\begin{figure}[t]
\includegraphics[width=14cm]{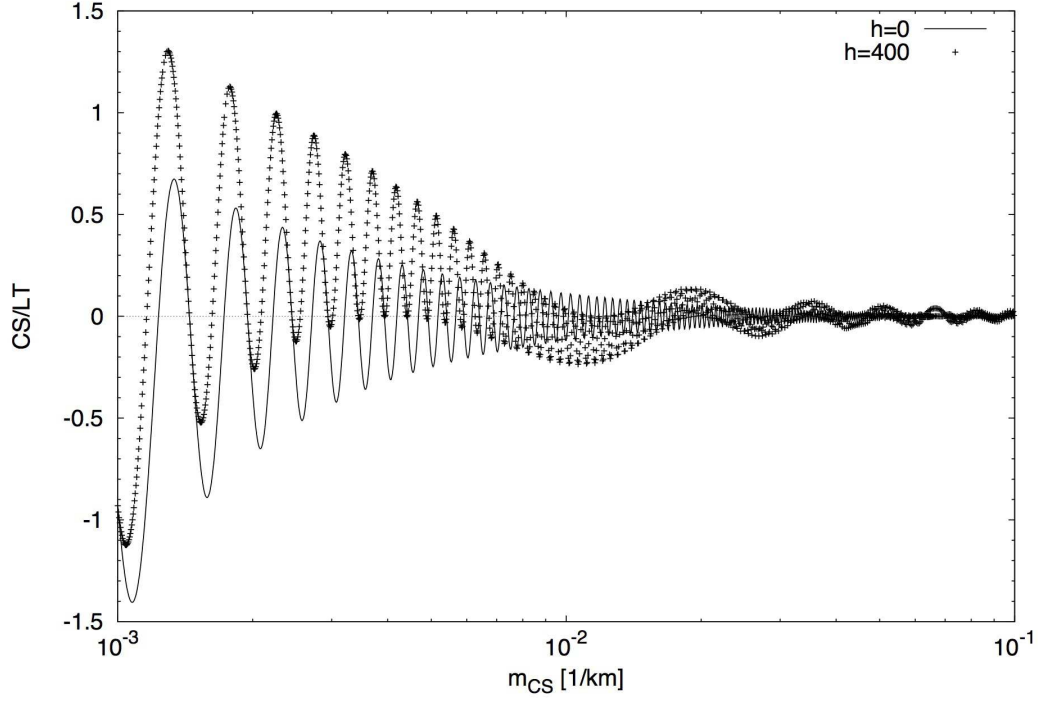}
\caption{ 
The ratio of $(c \Delta t)_{CS}/(c \Delta t)_{LT}$ at the ground level 
and the ISS site at $h \sim 400 \mbox{km}$. 
For its simplicity, we assume the equatorial case 
as $\phi = 0^{\circ}$ and the northbound direction as $\alpha = 0^{\circ}$. 
This figure suggests that the altitudinal effect might make 
a more complicated form of oscillating behavior in terms of $m_{CS}$ 
compared with the ground level. 
}
\label{f2}
\end{figure}
\begin{figure}[t]
\includegraphics[width=14cm]{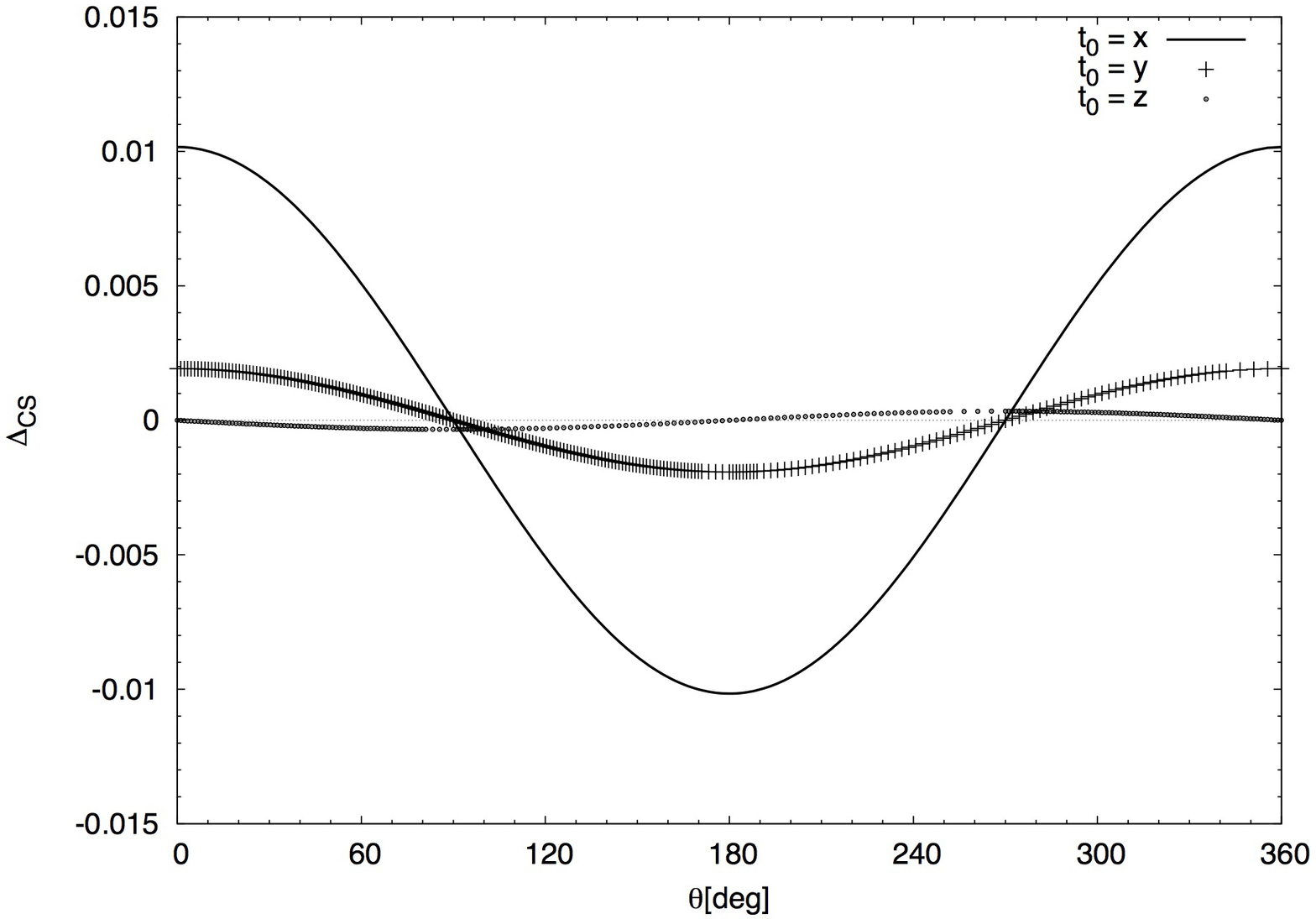}
\caption{ 
Time shift as a function of the orbital phase angle $\theta$ 
that is measured from the initial direction on the equator. 
For its simplicity, we assume the corotation of the ISS in a polar orbit 
around the Earth $(h \sim 400 \mbox{km})$. 
For the circular orbit, $\theta$ is proportional to time, 
where $\theta = 0^{\circ}$, $90^{\circ}$, $180^{\circ}$ and $270^{\circ}$ 
are corresponding to the passage time of the equatorial plane, 
a pole of Earth, the equatorial plane and the opposite pole, respectively. 
We consider three directions of the interferometer: 
The labels as x, y and z denote the eastbound direction, 
northbound one and vertical one, respectively, 
at the initial time as $\theta = 0^{\circ}$. 
}
\label{f3}
\end{figure}

Before closing this section, we mention 
the altitudinal dependence of the CS effect on the Sagnac interference. 
%
In order to understand the dependence, 
let us suppose two interferometers at different altitudes:  
One is located at $R$ and the other is at $R + h$. 
Eq. (\ref{deltat-CS}) suggests that a height difference $h$ 
makes a change in the time difference, 
where we assume the same interferometers. 
The relative difference between two measurements 
is of the order of $\sim m_{CS} h$. 
For instance, 
\begin{eqnarray}
\left|\frac{(c \Delta t)_{R + h} - (c \Delta t)_R}
{(c \Delta t)_R} \right|_{CS} 
&\sim&
\frac{h \times \frac{\partial}{\partial r} (c \Delta t)_R}
{(c \Delta t)_R}
\nonumber\\
&\sim& 
0.002 
\left(\frac{m_{CS}}{0.001 \mbox{km}^{-1}}\right)
\left(
\frac{h}{1600 \mbox{m}}  
\right) , 
\end{eqnarray}
where Eq. (\ref{deltat-CS}) is used and 
we assume Denver as a high city. 
Hence, the height difference of the CS effect is 
very small on the surface of the Earth. 
If the Sagnac interferometer were located 
in the space, however, 
the altitudinal difference might become significant as  
\begin{equation}
\left|\frac{(c \Delta t)_{R + h} - (c \Delta t)_R}
{(c \Delta t)_R} \right|_{CS}  
\sim 
0.4 
\left(\frac{m_{CS}}{0.001 \mbox{km}^{-1}}\right)
\left(
\frac{h}{400 \mbox{km}} 
\right) , 
\end{equation}
where we assumed the international space station (ISS) 
as an example. 
Figure \ref{f2} shows a comparison between the ground level and 
the ISS site regarding the oscillating behaviors in terms of $m_{CS}$.  
The altitudinal effect might make a more complicated form of 
oscillating behaviors. 
Such a altitudinal dependence might be helpful for a future test. 
Figure \ref{f3} shows the time shift as a function of 
the orbital phase angle $\theta$, which is nearly proportional to time 
because the ISS moves on a nearly polar orbit. 
The ISS is orbiting around the Earth with the period of nearly 90 minutes. 
Hence, the apparent zero-frequency strain due to the gravitomagnetic effects 
on such a satellite experiment can vary with time. 
This time variability might help separate the effects from the others. 
One could use this altitudinal dependence 
in order to place tighter constraints on the mass parameter 
in the future.  
For instance, a commercial ring laser as a long-term stable 
gyroscope is used in a number of airplanes. 
Such an instrument, if it is sufficiently improved in the future, 
may be used for the present purpose.


\section{Conclusion}
In this paper, we have investigated relativistic Sagnac effects 
in CS modified gravity. 
The altitudinal, latitudinal and directional dependence 
of relativistic Sagnac effect in the CS model 
is oscillatory in terms of the CS parameter $m_{CS}$. 

We have compared the CS effects on Sagnac interferometers with 
the general relativistic Lense-Thirring (LT) effects. 
LT effects on the eastbound interferometer cancel out. 
Therefore, our numerical calculations have suggested that 
the eastbound Sagnac interferometer 
might be preferred for testing CS separately. 

For some region of the CS parameter 
$m_{CS} \sim 0.01-0.001 [\mbox{km}^{-1}]$, 
the possible altitudinal dependence might become important 
when we consider a space experiment such as the ISS site 
at $h \sim 400 \mbox{km}$. 
Further investigations along this course might be interesting 
as a future work.

We would like to thank N. Yunes, S. Takeuchi and K. Edamatsu 
for useful discussions. 
We wish to thank the referee for useful comments on 
experimental realities and instabilities of nondynamical CS models. 
This work was supported in part 
by JSPS Grant-in-Aid for JSPS Fellows, No. 24108 (K.Y.), 
and JSPS Grant-in-Aid for Scientific Research (Kiban C), 
No. 26400262 (H.A.).

\begin{thebibliography}{99}
\bibitem{note}
Note that the nondynamical CS theory suffers from 
a rather serious instability \cite{Dyda} 
and hence it is not clear how viable this alternative theory is. 
In the future, the CS model adopted in the present paper 
might be replaced with a more improved one.
\end{thebibliography}
\end{document}